\documentclass[aps,prd,preprint,nofootinbib,superscriptaddress]{revtex4-1}
\usepackage{graphicx, epsfig}
\usepackage{color}
\usepackage{amsmath}
\usepackage{amssymb}
\newcommand{\be}{\begin{equation}}
\newcommand{\ee}{\end{equation}}
\newcommand{\bea}{\begin{eqnarray}}
\newcommand{\eea}{\end{eqnarray}}
\newcommand{\beal}{\begin{aligned}}
\newcommand{\eeal}{\end{aligned}}

\definecolor{dured}{RGB}{170,43,74}

\newcommand{\gapp}{\mathrel{\raise.3ex\hbox{$>$}\mkern-14mu
\lower0.6ex\hbox{$\sim$}}}
\newcommand{\lapp}{\mathrel{\raise.3ex\hbox{$<$}\mkern-14mu
\lower0.6ex\hbox{$\sim$}}}
\def\bbox{{\,\lower0.9pt\vbox{\hrule \hbox{\vrule height 0.2 cm
\hskip 0.2 cm \vrule  height 0.2 cm}\hrule}\,}}

\begin{document}

\preprint{DCPT-19/25}

\title{Connecting the Higgs Potential and Primordial Black Holes}
\author{De-Chang Dai}
\affiliation{Center for Gravity and Cosmology, School of Physics
Science and Technology, Yangzhou University, 180 Siwangting Road, 
Yangzhou City, Jiangsu Province, P.R. China 225002}
\affiliation{CERCA/Department of Physics/ISO, Case Western 
Reserve University, Cleveland OH 44106-7079}
\author{Ruth Gregory}
\affiliation{Centre for Particle Theory, Durham University,
South Road, Durham, DH1 3LE, UK}
\affiliation{Perimeter Institute, 31 Caroline St, Waterloo, Ontario N2L 2Y5,
Canada}
\author{Dejan Stojkovic}
\affiliation{HEPCOS, Department of Physics, SUNY at Buffalo, Buffalo, NY 14260-1500}


\begin{abstract}
\widetext
It was recently demonstrated that small small black holes can act 
as seeds for nucleating decay of the metastable Higgs vacuum, 
dramatically increasing the tunneling probability.  
Any primordial black hole lighter than $4.5 \times 10^{14}$g at formation
would have evaporated by now, and in the absence of new physics 
beyond the standard model, would therefore have entered the mass range 
in which seeded decay occurs, however, such true vacuum bubbles must percolate
in order to completely destroy the false vacuum; this depends on the bubble 
number density and the rate of expansion of the universe. 
Here, we compute the fraction of the universe that has decayed to the true vacuum
as a function of the formation temperature (or equivalently, mass) of the primordial
black holes, and the spectral index of the fluctuations responsible for their formation.
This allows us to constrain the mass spectrum of primordial black holes
given a particular Higgs potential and conversely, should we discover primordial 
black holes of definite mass, we can constrain the Higgs potential parameters.
\end{abstract}


\pacs{}
\maketitle

\section{Introduction}

One of the most fascinating implications of the measurement of the Higgs
mass at the LHC \cite{ATLAS:2012ae,Chatrchyan:2012tx} is that the 
standard model vacuum appears to be metastable 
\cite{Krive:1976sg,1982Natur.298,Sher:1988mj,
Isidori:2001bm,Degrassi:2012ry,Gorsky:2014una,Bezrukov:2014ina,
Ellis:2015dha,Blum:2015rpa}. Initially, this was not thought to be a 
problem for our universe, as standard techniques for computing vacuum decay 
\cite{Kobzarev:1974cp,coleman1977,callan1977,CDL} indicated that
the half-life was many order of magnitude greater the age of the universe.
However, vacuum decay represents a first order phase transition,
and in nature these typically proceed via catalysis: a seed or impurity
acts as a nucleus for a bubble of the new phase to form. In 
\cite{Gregory:2013hja,Burda:2015isa,Burda:2015yfa,Burda:2016mou,
Cuspinera:2018woe}, the notion that a
black hole could act as such a seed was explored, with the 
finding that black holes can dramatically shorten the lifetime of a
metastable vacuum (see also \cite{PhysRevD.35.1161,Berezin:1987ea,
Tetradis:2016vqb,Chen:2017suz,Gorbunov:2017fhq,
Mukaida:2017bgd,Oshita:2018ptr,Kohri:2017ybt}). 
Interestingly, before the discovery of the Higgs particle, the 
electroweak phase transition was usually described as a
second order transition, and in \cite{Greenwood:2008qp} the idea that
the usual second order electroweak phase transition might be followed 
by a first order phase transition was explored.

For a black hole to seed vacuum decay, we must be sure that the half-life
for decay is less than the evaporation rate of the black hole. This means
that the branching ratio of tunnelling to decay must be greater than one.
In \cite{Burda:2015yfa,Burda:2016mou} this was found to occur for
black holes of order $10^{6-9}M_p$ or so, by which point
the half-life for decay is of order $10^{-23}$s! Clearly this process is 
not relevant for astrophysical black holes, however, it has been hypothesised
that there exist very light black holes, formed from extreme density
fluctuations in the early universe \cite{Hawking:1971ei,Carr:1974nx,
Khlopov:2008qy} dubbed {\it primordial black holes}. Such black holes
have been proposed as a source for dark matter \cite{Carr:2016drx},
and although this has now been ruled out \cite{Zumalacarregui:2017qqd},
they could still constitute a component of the dark matter of the universe. 
Indeed, it has even been proposed that the Higgs vacuum instability could 
generate primordial black holes in the early universe \cite{Espinosa:2017sgp}.

Given that we are in a current metastable Higgs vacuum, we can be sure that there
has been no primordial black hole that has evaporated in our past lightcone,
however, how strong a constraint on primordial black holes can we place?
For the universe to have decayed, the black hole must not only have 
evaporated sufficiently to reach the mass range in which catalysis 
spectacularly dominates, but the consequent bubble (or bubbles)
of true vacuum must have percolated to engulf the current Hubble volume. 
Thus, this is a statement
about the relative volume in the percolated bubble, which is itself a
statement on the primordial black hole density and mass. In this paper,
we draw together all these aspects of the problem, linking the primordial
black hole spectral index and formation epoch to the standard model
parameters. 

The outline of the paper is as follows.
In section \ref{hvd} we review the physics of the Higgs vacuum decay 
in the presence of gravity. In section \ref{bhm} we relate the primordial 
black hole masses that can trigger vacuum decay with the 
parameters in the effective Higgs potential. In section \ref{cos} we 
put this scenario in the cosmological context: Every black hole that 
can trigger the vacuum decay will create a bubble of true vacuum. 
These bubbles then expand with the 
speed of light, but their number density decreases due to the 
expansion of the universe. For a successful phase transition, 
the bubbles have to percolate, so we define a quantity ${\cal P}$, which 
represents the portion of the universe that has already transitioned to the 
new vacuum. For ${\cal P} \geq 1$, the universe would be destroyed, thus the 
associated range of parameters is excluded. We summarise and discuss 
our findings in section \ref{con}.

\section {False vacuum decay with black holes}
\label{hvd}

The high energy effective Higgs potential has been determined by a 
two-loop calculation in the standard model as \cite{Degrassi:2012ry}
\begin{equation}
V(\phi )= \frac{1}{4}\lambda_{\rm eff} (\phi) \phi^4  ,
\end{equation}
where $\lambda_{\rm eff}(\phi)$ is the effective coupling constant that
runs with scale.
We now review the calculations in \cite{Burda:2016mou}, adopting the same
conventions. The running of the coupling constant can be excellently modelled 
over a large range of scales by the three parameter fit:
\begin{equation}
\lambda_{\rm eff} =\lambda_* +b\Big( \ln\frac{\phi}{M_p}\Big)^2 
+c \Big( \ln\frac{\phi}{M_p}\Big)^4 ,
\label{lambdaeff}
\end{equation}
where $M_p^{-2} = {8\pi G}$. By fitting the two-loop calculation with
a simple analytic form, we can easily investigate not only the standard
model, but also beyond the standard model potentials, allowing us to
explore possible future corrections to the standard model results.

The Higgs potential supports a first 
order phase transition mediated via nucleation of bubbles of new 
vacuum inside the old, false, vacuum.  The nucleation rate in the presence of 
gravity is determined by a saddle point `bounce' solution of the Euclidean 
(signature $+,+,+,+$) action:
\begin{equation}
S_E = \int_\mathcal{M}\left[-\frac{1}{16\pi G} \mathcal{R}
+\left(\frac{1}{2}g^{ab}\partial_a \phi \partial_b \phi +V(\phi) \right)\right]
\sqrt{-g}\,d^4x  .
\end{equation}
The spacetime geometry is taken to have $SO(3)\times U(1)$ symmetry,
in other words, it is spherically symmetric ``around'' the black hole, and has
time translation symmetry along the Euclidean time direction, $\tau$:
\begin{equation}
ds^2 = f(r) e^{2\delta(r)} d\tau^2 +\frac{dr^2}{f(r)}
+r^2 (d\theta^2 +\sin^2\theta d\varphi^2)\,,
\end{equation}
with
\begin{equation}
f(r)=1-\frac{2G\mu (r)}{r}\,.
\end{equation}
We can think of $\mu(r)$ as the local mass parameter, however caution 
must be used in pushing this analogy. For an asymptotic vacuum of $\Lambda=0$,
then $\mu(\infty)$ truly is the ADM mass of the black hole, however, locally, 
$\mu$ also includes the effect of any vacuum energy: for a pure Schwarzschild-(A)dS
solution, $\mu(r) = M + \Lambda r^3/6G$. Since we are interested in seeding
the decay of our current SM vacuum, we will take $\Lambda_+=0$, so that the
asymptotic value of $\mu$ is indeed the seed black hole mass, $M_+$, 
responsible for triggering the phase transition. The remnant mass, 
which is a leftover from the seed black hole after some of its energy is 
invested into the bubble formation, may not be precisely $\mu(r_h)$, however,
since we will be interested only in the {\it area} of the remnant black hole 
horizon, it turns out that $\mu(r_h)$ is in fact the desired quantity.

The Higgs and gravitational field equations of motion are
\be
\beal
&f\phi''+f'\phi'+\frac{2}{r}f\phi'+\delta'f\phi'-V_{,\phi } =0\\
&\mu' =4\pi r^2 \Big(\frac{1}{2}f{\phi'}^2+V\Big)\\
&\delta '  =4\pi Gr{\phi'}^2 ,
\eeal
\ee
where $V_{,\phi } \equiv \partial V/\partial \phi$.
The black hole horizon is at $r=r_h$, at which $f(r_h)=0$. 
We have to solve these equations of motion numerically in order to get the 
function $\phi (r)$, and to do this, we start from the horizon $r_h$
with a particular remnant parameter, $r_h=2G\mu_-$, and some 
value for the Higgs field $\phi_h$. At the horizon therefore
the fields satisfy the boundary conditions
\be
\beal
\mu(r_h)&=\mu_- \;\; \text{, } \qquad \delta(r_h)=0 \\
\phi'(r_h)&=\frac{r_h V_{,\phi }(\phi_h)}{1-8\pi G r_h^2 V(\phi_h)} \,,
\eeal
\ee
and as $r\rightarrow \infty$, 
\be
\lim_{r \rightarrow \infty} \phi(r)\rightarrow 0  \quad, \qquad 
M_+=\lim_{r\rightarrow \infty}\mu(r) .
\label{infbc}
\ee
We use a shooting method starting at $r_h$ with $\phi=\phi_h$ and integrate
out, altering $\phi_h$ until a solution is obtained with $\phi$ tending
to $0$ for very large values of $r$. In practise, rather than setting the
asymptotic mass $\mu(\infty) = M_+$, we set the initial (remnant) value of 
$\mu_-$ and deduce the seed mass from \eqref{infbc}, repeating the integration 
for a range of values of $\mu_-$. We then determine $\mu_-(M_+)$ by
inversion. 

The decay rate of the Higgs vacuum, $\Gamma_D$, is then determined by 
computing the difference in entropy between the seed and remnant black
holes: 
\be
\Gamma_D=\Big(\frac{B}{2\pi}\Big)^{1/2}(GM_+)^{-1}e^{-B}
\ee
where
\be
B=\frac{M_+^2-\mu_-^2}{2M_p^2} .
\ee
As pointed out in \cite{Burda:2015isa,Burda:2015yfa,Burda:2016mou}, a black 
hole can also radiate and lose mass, eventually disappearing in Hawking 
radiation, at a rate initially estimated by Page \cite{Page1976}, see also 
\cite{Carr:1994ar,Carr:1998fw,Carr:2009jm,MacGibbon:2015mya,Carr:2016hva}:
\begin{equation}
\Gamma_H=3.6\times 10^{-4} (G^2M_+^3)^{-1} .
\end{equation}
Thus, we define the branching ratio between the tunneling and evaporation rate as
\begin{equation}
\frac{\Gamma_D}{\Gamma_H}=43.8 \frac{M_+^2}{M_p^2}B^{1/2}e^{-B} .
\label{br}
\end{equation}
This equation contains all the information we need. In the next two sections, 
we will study the consequences of the gravitationally induced false Higgs 
vacuum decay.

\section {The vacuum decay rate and the Higgs effective potential }
\label{bhm}

If the branching ratio given by Eq.~(\ref{br}) is larger than one, then the 
tunneling rate is faster than evaporation rate, and the black hole can 
catalyze false vacuum decay. 
Note that the branching ratio depends on three parameters: 
$M_+$, $\lambda_*$, and $b$: fitting the form of $\lambda_{\rm eff}$
in \eqref{lambdaeff}
to the standard model value at the electroweak scale fixes $c$ in
terms of $\lambda_*$ and $b$, and
$M_+$ is the primordial black hole seed.

\begin{figure}
\centering
\includegraphics[width=12cm]{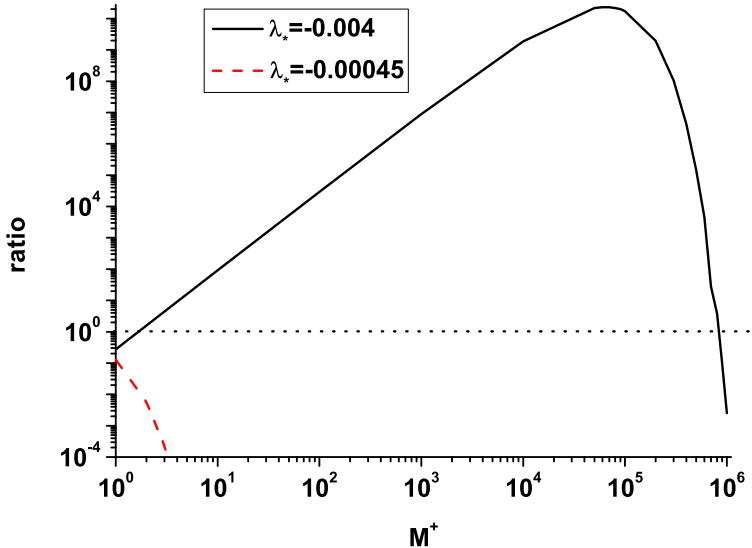}
\caption{The  branching ratio between the tunneling and evaporation rate 
as a function of the seed black hole mass, $M_+$, for some sample choices 
of the potential parameters. The unit branching ratio, 
$\Gamma_D/\Gamma_H=1$, is labeled by the dotted line. 
We set $b=1.5\times 10^ {-5}$, $c=0$. $M_+$ is given in units of $M_p$.
}
\label{compare}
\end{figure}
Let us first illustrate the results for some sample choices of the potential parameters.
If we set $\lambda_*=-0.004$, $b=1.5\times 10^ {-5}$, $c=0$, then 
Fig.~\ref{compare} shows that the branching ratio is larger than one for
\begin{equation} \label{bhmr}
M_p \lesssim M_+\lesssim 10^6 M_p.
\end{equation}
This means that primordial black holes with masses within this range 
can initiate Higgs vacuum decay for the associated values of the 
Higgs potential parameters. A black hole mass with the lifetime of the 
current age of the universe is approximately $ 4.5 \times 10^{14}$ grams,
meaning that all black holes lighter than this value would have already
evaporated. Along the way, they will inevitably end up in the range given by 
Eq.~(\ref{bhmr}). This however does not automatically imply that all the 
primordial black holes lighter than $ 4.5 \times 10^{14}$ grams are 
excluded for this choice of parameters. To destroy the universe the 
bubbles of the true vacuum have to percolate, which takes time. 
We will study this in the next section.

The same Fig.~\ref{compare} indicates that if we set $\lambda_*=-0.00045$, 
and keep $b=1.5\times 10^ {-5}$ and $c=0$, then the branching ratio is 
always smaller than one (these values are not consistent with a pure 
standard model effective coupling, however, indicate the principle of
model dependence of the branching ratio). 
In that case, the primordial black holes of any 
mass (i.e.  $M_p< M_+ < \infty $) cannot stimulate the false vacuum to 
decay into true vacuum, and our universe is safe. We excluded the black 
hole seed masses less than $M_p$ from the discussion, as the semi-classical
approximation used in computing the decay rate is no longer expected to
be valid at the Planck scale, where presumably a full theory of quantum gravity
is required.

\begin{figure}
\centering
\includegraphics[width=12cm]{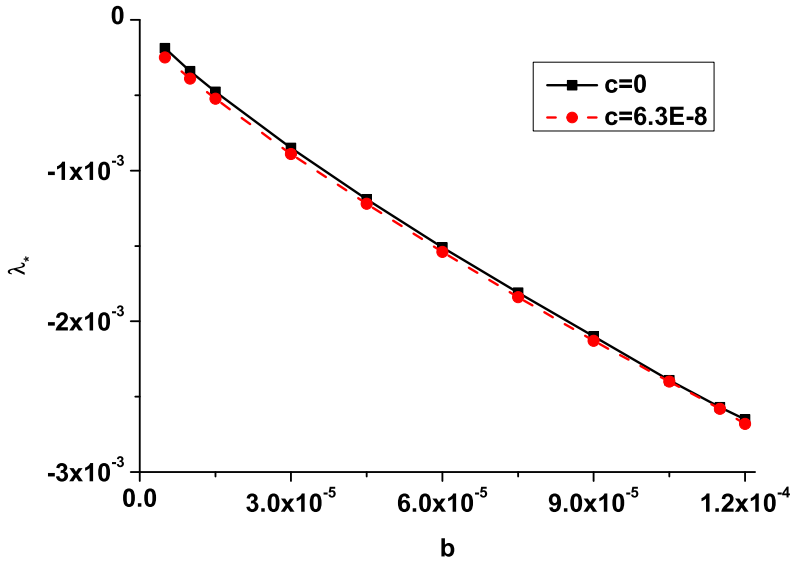}
\caption{A plot of the line $\frac{\Gamma_D}{\Gamma_H}=1$ in 
$\left ( b, \lambda_*\right)$ parameter space for the values
$c=0, 6.3\times 10^{-8}$ as labelled. Above the 
curve, the branching ratio $\frac{\Gamma_D}{\Gamma_H}$ is always 
less than one for any value of the seed black hole mass. Below the 
curve, there is always a range of the black hole masses for which the 
branching ratio $\frac{\Gamma_D}{\Gamma_H}$ is greater than one. 
The dependence on $c$ is minimal.
}
\label{boundary}
\end{figure}

It is now instructive to systematically analyze the range of parameters 
for the effective coupling \eqref{lambdaeff}. Fig.~\ref{boundary} shows 
the threshold curve $\frac{\Gamma_D}{\Gamma_H}=1$ in $\left ( b,
\lambda_*\right)$ parameter space for two values of the parameter $c$.  
The region of parameter space with $\frac{\Gamma_D}{\Gamma_H}<1$, 
for which the universe is safe, is above the curve. Below the curve, 
the branching ration will be greater than one for some range of black 
hole masses (similar to that shown in Eq.~(\ref{bhmr})) below
the quantum gravity scale. This range is different for differing $\lambda_*$, 
$b$, and $c$ (so not easy to plot) however, it can easily calculated by 
substituting the concrete values for  $\lambda_*$, $b$, and $c$, in Eq.~(\ref{br}).
The boundary with $c=6.3\times 10^{-8}$ is lower than that with $c=0$ 
because of the contribution from the quartic terms in the Higgs potential. 
However, numerical experiments indicate that the curves do not change 
significantly as we vary the parameter $c$. 

According to \cite{Burda:2016mou} the standard model parameter space 
corresponding to the allowed range of top quark
mass ($172$ - $174$ GeV) is 
$1.2\times 10^{-5}\lesssim b \lesssim 1.4\times 10^{-5}$ and 
$-0.02\lesssim \lambda_* \lesssim -0.007$. Unfortunately, this lies 
inside the potentially dangerous $\frac{\Gamma_D}{\Gamma_H}>1$ region!
If some new physics beyond the standard model modifies the Higgs 
potential, then we can potentially move outside of these parameter ranges. 
In the best case, the Higgs vacuum gets stabilized and the first order 
phase transition disappears altogether from the discussion. 
However, there is also a potentially dangerous region where 
beyond the standard model corrections give a `safe' long false 
vacuum lifetime in the absence of black holes, yet unacceptably short 
in their presence. Thus, one can always tension new 
beyond the standard model physics with the existence of the primordial 
black holes of a certain mass.

\section{Primordial black hole masses and percolating bubbles}
\label{cos}

In the previous section, we saw that any primordial black hole that had 
enough time to evaporate sufficiently to fit into an appropriate mass range 
for the corresponding choice of the parameters $\lambda_*$ and $b$, 
could initiate false vacuum decay. The bubbles of true 
vacuum then expand with the speed of light, but the background universe 
expands as well. Successful completion of the first order phase transition 
depends on the number density of the created bubbles. In our scenario, 
every black hole that can initiate the false vacuum decay will create a 
bubble, so the number of the bubbles is equal to the number of such 
primordial black holes. Thus, whether the initiated vacuum decay can 
be completed crucially depends on the production mechanism and age 
of the universe when the primordial black holes were formed.

It is usually assumed that primordial black holes are produced by 
density fluctuations caused by oscillations of some (scalar) field. 
If density fluctuations are large enough 
\cite{Green:1997sz,Kodama:1982sf,Crawford:1982yz,
Hsu:1990fg,Sanz:1985np,Harada:2013epa}, 
the whole causally connected region (i.e.\ the horizon volume at some time) 
collapses and forms a black hole.
The horizon mass in a radiation-dominated universe (in units of grams) is
\begin{equation}
\label{eqn:Mass}
M_H \cong 10^{18} \left(\frac{10^7 GeV}{T}\right)^2\,{\rm g}\,.
\end{equation}
where $T$ is the temperature of radiation. Obviously, the earlier  
the black holes are formed, the lighter they are, hence their lifetime is shorter. 
Their lifetime is given as \cite{Page1976,Carr:1994ar,Carr:1998fw,
Carr:2009jm,MacGibbon:2015mya,Carr:2016hva}.
\begin{equation} \label{evap}
\tau_{\rm evap} = 4.99 \times 10^{-44}\Big(\frac{ M}{M_{P}}\Big)^3\,{\rm s}\,.
\end{equation}
Black holes of mass $M \gtrsim 4.5 \times 10^{14}\,$g have a 
lifetime greater than $1.38\times 10^{10}$ years, or the age of the universe.  
Therefore only lighter primordial black holes will have the potential to 
destroy the universe, based on the SM parameter ranges. 
We focus mostly on these lighter black holes which, 
according to Eq.~(\ref{eqn:Mass}), are created at temperatures higher 
than  $T_F\gtrsim 4.7\times 10^{8}$ GeV.

After primordial black holes are formed at  
$T_F$, their number density changes with temperature as
\begin{eqnarray}
\label{eqn:density-1}
n_b(T)&\sim &\frac{\beta_{\rm i}}{M_{i}}\, \rho_{\rm r,i}
\left( \frac{T}{T_F}\right)^3.
\end{eqnarray}
where $\beta_{\rm i}$ is the mass fraction of the universe in black holes 
at formation, while  $\rho_{\rm r,i}=\frac{\pi^2}{30}g_F T^4_F$ 
is the radiation energy density at that time, with $g_F \approx 100$ 
being the number of degrees of freedom of radiation species at $T_F$. 
$M_F$ is the mass of the primordial black holes at formation, and  
we take $M_F =M_H(T_F)$ as usual. The mass fraction $\beta_{i}$ can be 
found assuming a Gaussian perturbation spectrum of fluctuations that 
lead to black hole formation (see e.g.\ \cite{Green:1997sz,Liddle:1993fq,
Bringmann:2001yp})
\begin{equation}
\label{eqn:b2}
\beta_i \approx \frac{\sigma_H (T_F)}{\sqrt{2\pi}\delta_{\rm min}}
e^{-\frac{\delta_{\rm min}^2}{2\sigma^2_H (T_F)}} ,
\end{equation}
The parameter $\delta_{\rm min}\approx 0.3$ is the minimum density contrast 
required for black hole creation, while $\sigma_H (T)$ is the mass variance 
evaluated at horizon crossing at the temperature $T$ defined as \cite{Liddle:1993fq}
\begin{equation}
\label{eqn:fluc}
\sigma_H (T_F) = \sigma_H (T_0) 
\left( \frac{M_H (T_0)}{M_H (T_{\rm eq})}\right)^{\frac{n-1}{6}}
\left( \frac{M_H(T_{\rm eq})}{M_H(T_F)}\right) ^{\frac{n-1}{4}} .
\end{equation}
Here, $T_{\rm eq}\approx 0.79eV$ is the temperature at the matter/radiation 
equilibrium, $T_0=2.725K=2.35\times 10^{-4}eV$ is the present temperature 
of the universe, while $n$ is the spectral index of the fluctuations that lead to 
black hole formation, i.e.\ $P(k)\propto k^n$. Note, the cosmic microwave 
background data indicate that the value of the spectral index of the {\it inflaton}
field is $n\approx 1$, however the CMB data probe the scales between 
$10^{45}$ and $10^{60}$ times larger than those probed by primordial 
black holes. It is expected that primordial black holes are formed by 
fluctuations of fields other than the inflaton (e.g.\ during phase transitions),
and the typical value of $n$ used in this context is between $1.23$ and $1.31$ 
\cite{Green:1997sz,Bringmann:2001yp,Kim:1999iv}. To normalize 
Eq.~(\ref{eqn:fluc}) we use the mass variance evaluated at the horizon 
crossing $\sigma_H(T_0)=9.5\times 10^{-5}$.
\begin{figure}
\centering
\includegraphics[width=10cm]{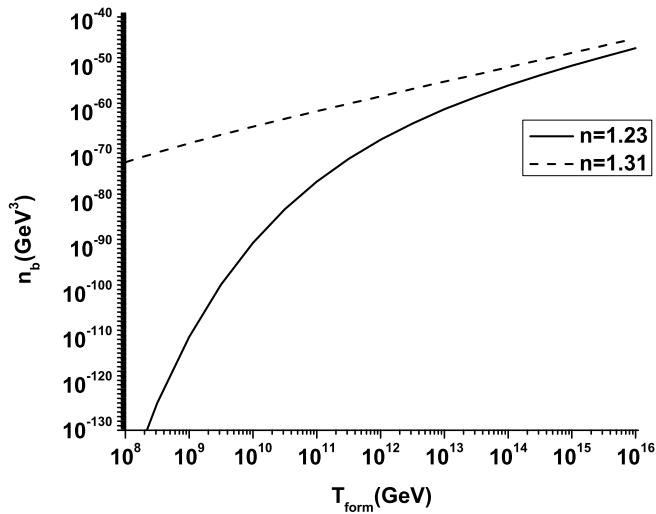}
\caption{The present number density of the true vacuum bubbles, 
plotted as a function of the temperature at which the primordial black holes
that seed nucleation are formed. The number density is shown for two
values of the spectral density index, $n$, of the perturbations 
responsible for the primordial black hole creation. 
}
\label{density}
\end{figure}

We now have all the elements to calculate the black hole abundance 
for any set of desired parameters.
After formation, primordial black holes evaporate, and at some stage of 
their life they will trigger false vacuum decay. When exactly this will 
happen depends on the specific parameters of the Higgs potential; we
must be above the threshold value of the branching ratio, or in the
range of parameters below the curve in Fig.~\ref{boundary}, where 
it is guaranteed that the phase transition will be initiated for 
some black hole mass range. 

To illustrate the procedure, we calculate the excluded primordial black 
hole parameter space for the example from Section \ref{hvd}, i.e.\ for the 
values of the potential parameters $\lambda_*=-0.004$, $b=1.5\times 10^ {-5}$, 
$c=0$. As shown in section \ref{hvd}, the branching ratio is larger than 
one for the seed black hole masses $M_p \lesssim M_+\lesssim 10^6 M_p$,
therefore all the black holes that have evaporated down to $10^6 M_p$ or 
less by the present time will trigger false vacuum decay for this set of 
parameters. We note that this number is effectively the same as the number 
of the black holes that have evaporated completely by the present time, 
since it takes only a fraction of the second for a black hole to evaporate 
from $10^6 M_p$ to zero. For comparison, we will also add black 
hole masses that correspond to the parameters beyond the standard model.

The scenario is as follows. Suppose that primordial black holes are
formed at a temperature $T_F$ with some initial mass $M_i$. They then 
evaporate until they reach a mass at which vacuum decay is catalysed
($10^{6-9} M_p$ for the standard model values, which is essentially equivalent 
to a complete evaporation, given the scale of the lifetimes involved).
For the parameters of the Higgs potential outside of the standard model, 
these masses could be much higher. At that moment 
(which depends on the initial black hole mass) they seed vacuum decay 
and form a bubble of true vacuum that then expands at the speed of light.
For a successful phase transition, the bubbles have to percolate, so we compute
the overall volume of true vacuum in the expanding universe from the volume
of an individual bubble and the number density of black holes.

The present time number density of the bubbles, $n_{\rm b}(T_0)$,  
is shown in Fig.~\ref{density}. It is calculated from Eq.~(\ref{eqn:density-1}) 
following the procedure outlined above. The present time radius of the bubble 
depends on the time it was created.
If an object (in this case a bubble of true vacuum) is created at a 
cosmological redshift $Z$, its present age, $t$, is given by
\be
t(Z) = \frac{1}{H_0}\int_{0}^Z \frac{dz}{E(1+z)}
\ee
where
\be
E^2 = \frac{H^2}{H_0^2} = \Omega_m (1+Z)^3+\Omega_{\rm rad}(1+Z)^4
+\Omega_k (1+Z)^2 +\Omega_\Lambda  \label{E}
\ee
Here $\Omega_m$, $\Omega_{\rm rad}$, $\Omega_k$ and 
$\Omega_\Lambda$ are the present values of the dark matter, radiation, 
curvature, and dark energy density respectively. We take their numerical 
values from Planck results \cite{Aghanim:2018eyx}, $\Omega_m=0.315$, 
$\Omega_{\rm rad} = 9.23 \times 10^{-5}$, $\Omega_k=0$ and 
$\Omega_\Lambda =0.684$. $H_0$ is the present time Hubble 
parameter, $H_0=67.4$ km s$^{-1}$ Mpc$^{-1}$. The connection 
between the temperature, $T$, scale factor, $a$, and the redshift, 
$Z$, in an expanding Friedman Robertson Walker universe is 
$T \propto 1/a \propto (1+Z)$.

Since $dr =c dt/a = cdt (1+Z)$, the current physical radius of the true vacuum 
bubble formed at redshift $Z_B$ is
\begin{equation}
R =\frac{c}{H_0}\int_{0}^{Z_B} \frac{dz}{E}
\end{equation}
where $E$ is given by Eq.~(\ref{E}). The redshift, $Z_B$, is calculated 
at the moment when a black hole of a certain seed mass (formed at the 
temperature $T_F$) evaporates enough to fit into the appropriate 
mass window where it can trigger the false vacuum decay.

Thus, the portion of the universe which is already in the new vacuum 
at the present time is
\begin{equation}
{\cal P}=\frac{4\pi}{3} R^3 n_{\rm b}(T_0) ,
\end{equation}
\begin{figure}
\centering
\includegraphics[width=10cm]{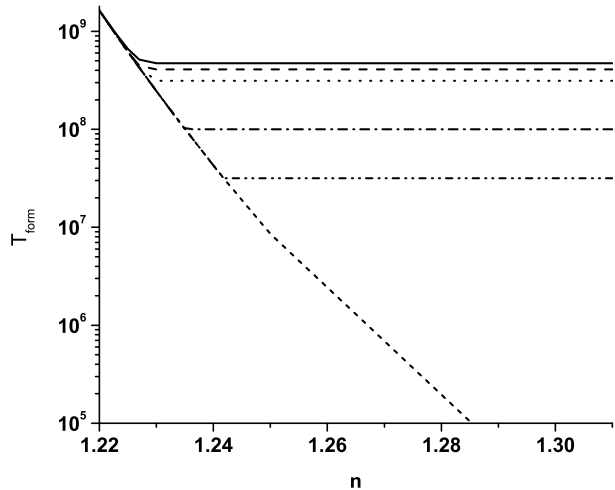}
\caption{This plot shows the temperature of the universe at the time 
of the primordial black hole formation, $T_F$, as a function 
of the spectral index, $n$ for several values of the black hole masses that 
can trigger the vacuum decay.  The lines represent the ${\cal P}=1$ value 
for the portion of the universe which already transitioned to the new 
vacuum at present time. Above these lines, we have ${\cal P}>1$ and 
the whole universe today would be destroyed. Below these lines we 
have ${\cal P}<1$, and the universe is safe.
A black hole of the initial mass $M_i$ (determined by the formation 
temperature $T_F$) evaporates and triggers the vacuum decay at 
some lower value $M_{\rm dec}$. 
If $M_i >M_{\rm dec}$, the onset of the phase transition is delayed by 
the time needed for a black hole to enter the window where it can trigger 
the vacuum decay with $\frac{\Gamma_D}{\Gamma_H}>1$. 
The solid line corresponds to the black hole masses 
$M_{\rm dec} \ll 5\times 10^{14}$g, i.e. $M_{\rm dec} \approx 0$.  
The dashed, dot, dashed dot, and dashed dot dot lines correspond to 
$M_{\rm dec}$ values  of $5\times 10^{14}$ g, $ 10^{15}$g, $ 10^{16}$g, 
$10^{17}$g. The short dashed line corresponds to $M_i<M_{\rm dec}$, 
where such black holes trigger the vacuum decay immediately upon formation.
}
\label{pro}
\end{figure}

Fig.~\ref{pro} shows the boundary of the ${\cal P=}1$ region. For the 
range of parameters in the upper part of the plot the universe today is 
destroyed, since the bubbles percolate. In contrast, for the range of 
parameters in the lower part of the plot, the universe is safe, though 
the primordial black holes may initiate false vacuum decay.
We distinguish between the initial black hole mass $M_i$ (determined by 
the formation temperature $T_F$) which evaporates and triggers the vacuum 
decay at some lower value $M_{\rm dec}$, determined by the 
requirement $\frac{\Gamma_D}{\Gamma_H}>1$. To show how the 
results vary with the mass of the black hole that triggers the vacuum 
decay, we plot the curves corresponding to several values of $M_{\rm dec}$.

\begin{figure}
\centering
\includegraphics[width=10cm]{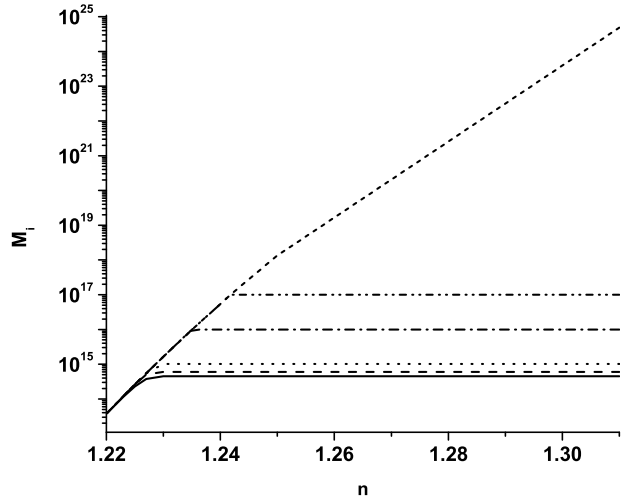}
\caption{This plot shows the the primordial black hole masses (in grams) 
at the time of their formation, $M_i$, as a function of the spectral index, $n$.  
 The lines represent the ${\cal P}=1$ value 
for the portion of the universe which already transitioned to the new 
vacuum at present time. Below this line we have 
${\cal P}>1$ and the whole universe today would be destroyed. 
Above this line, we have ${\cal P}<1$, and the universe is safe.
The solid line corresponds to the black hole masses 
$M_{\rm dec} \ll 5\times 10^{14}$g, i.e. $M_{\rm dec}\approx 0$.  
The dashed, dot, dashed dot, and dashed dot dot lines correspond 
to $M_{\rm dec}$ values  of $5\times 10^{14}$ g, $ 10^{15}$g, $ 10^{16}$g,
$10^{17}$g. The short dashed line corresponds to $M_i<M_{\rm dec}$, 
where such black holes trigger the vacuum decay immediately upon formation.
}
\label{prom}
\end{figure}

\begin{figure}
\centering
\includegraphics[width=10cm]{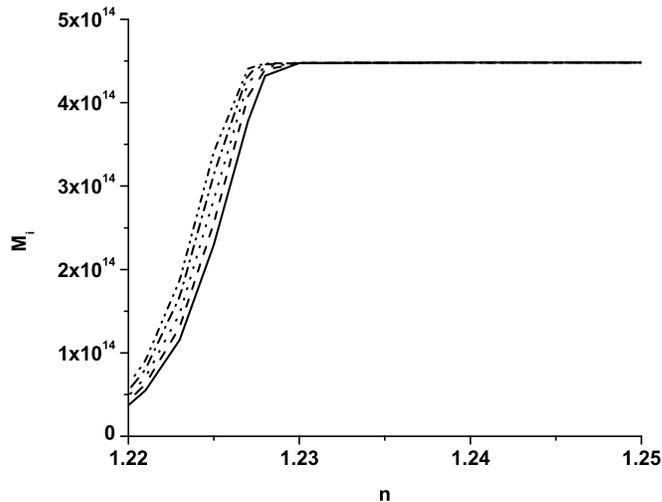}
\caption{This plot shows the the primordial black hole masses (in grams) 
at the time of their formation, $M_i$, as a function of the spectral index, $n$.  
The solid lines represent the ${\cal P}=1$ value for the portion of the universe 
which already transitioned to the new vacuum at present time. Below this line 
we have ${\cal P}>1$ and the whole universe today would be destroyed. 
Above this line, we have ${\cal P}<1$, and the universe is safe. The dashed, 
dot, dashed dot and dashed dot dot  lines represent the values ${\cal P}=10^{-1}$, 
${\cal P}=10^{-2}$, ${\cal P}=10^{-3}$ and ${\cal P}=10^{-4}$ respectively, where 
the phase transition has not completed yet, but there are bubbles of true 
vacuum present in the universe.}
\label{prom-p}
\end{figure}

With the help of Eq.~(\ref{eqn:Mass}), we can convert the temperature 
of the universe at the time of the primordial black hole formation to the 
primordial black hole mass. This is shown in Fig.~\ref{prom}. 
We can see that lighter black holes are more dangerous than the 
more massive ones because they evaporate quickly, form the true 
vacuum bubbles earlier, and the bubbles have more time to grow. 
Black holes much heavier than $M \gtrsim 10^{17} g$ 
are not constrained since they did not have enough time to 
evaporate down to the dangerous range of masses. 
A potentially interesting range of the parameter space is where ${\cal P}$ 
is nonzero but not too close to one. This would mean that there are a 
few bubbles in the universe here and there, but they do not yet dominate 
the universe. This situation is shown in Fig. \ref{prom-p}.  Since the bubbles expand with the speed of 
light, if they are far from us, this state could last for billions of years.  
It would be very interesting to study observational effects of the 
presence of such bubbles within our horizon.

\section{Conclusions}
\label{con}
We demonstrated here that it is possible to connect the parameters of 
the Higgs potential with the primordial black hole masses and physics 
of their formation (in our case the spectral index of perturbations that 
leads to their formation). We used the recent result that corrections 
due to black hole seeds can significantly increase the tunneling probability from 
the false to true Higgs vacuum. Any primordial black hole that had 
enough time to lose its mass from its formation till today to fit into 
an appropriate mass range  for the corresponding choice of the 
Higgs potential parameters could trigger the false Higgs vacuum 
decay. If there is no new physics beyond the standard model, any black hole
lighter than $4.5 \times 10^{14} g$ could trigger the decay. If some 
new physics beyond the standard model modifies the Higgs potential, 
then the black hole masses change appropriately. The decay
rate is proportional to the exponential of the difference in entropy
of the seed and remnant black holes masses, roughly $(M_++\mu_-)(M_+-\mu_-)$.
Numerically, this exponent varies slowly with $M_+$, and so can be
regarded as $M\delta M$. The branching ratio that determines when
the decay dominates evaporation is therefore most sensitive to the
difference in seed and remnant masses that in turn is determined by
an integral of the energy momentum of the scalar field bounce solution.
Therefore, any potential that has a very thick bounce solution and correspondingly
small $\delta M$ will have a much higher threshold of black hole mass
for vacuum decay catalysis. We explored the threshold
for a range of masses including the primordial black holes that could
contribute to dark matter. Such black holes are outside the current SM 
parameter ranges, but could potentially be relevant for BSM potentials.

However, just triggering the decay is not enough to destroy the 
universe, and automatically exclude associated black hole range.
For a successful completion of the first order phase transition the 
bubbles have to percolate, which in turn depends on the number 
density of the created bubbles. Since every black hole that can 
initiate the false vacuum decay will create a bubble, the number 
of the bubbles is equal to the number  of such primordial black holes. 
We then trace evolution of the bubbles. The bubbles of the true 
vacuum expand with the speed of light, but the background universe 
expands as well, so their number density decreases.
We define a quantity ${\cal P}$, which represents a portion of the universe 
which already transitioned to the new vacuum at the current time. 
For ${\cal P} \geq 1$, the universe is destroyed, and the associated range 
of parameters is excluded.

Our procedure can be used in two ways. If we use the Higgs potential 
parameters as an input, we can constrain the black hole masses and 
the physics of formation (e.g. the spectral index of perturbations). 
In turn, if we ever discover primordial black holes of definite mass, 
we can use it to constrain the Higgs potential parameters, or indeed
the presence of extra dimensions \cite{Mack:2018fny,Cuspinera:2019jwt}.

\begin{acknowledgments}
It is a pleasure to thank Ian Moss for helpful discussions.
D.C Dai was supported by the National Science Foundation of China 
(Grants 11433001 and 11775140), National Basic Research Program of 
China (973 Program 2015 CB857001),  and  the Program of Shanghai 
Academic/Technology Research Leader under Grant No.\ 16XD1401600. 
RG is supported in part by the Leverhulme grant \emph{Challenging the 
Standard Model with Black Holes}, in part by STFC consolidated grant 
ST/P000371/1, and in part by the Perimeter Institute.
D.S. was partially supported by the US National Science Foundation, under 
Grant No. PHY-1820738.
Research at Perimeter Institute is supported by the Government of
Canada through the Department of Innovation, Science and Economic 
Development Canada and by the Province of Ontario through the
Ministry of Research, Innovation and Science.

\end{acknowledgments}

\end{document}